\documentclass[prl,amsmath,twocolumn,showpacs]{revtex4-1}

\usepackage{epsfig}
\usepackage{amsfonts} 
\usepackage{nicefrac}
\usepackage{color} 
\usepackage{hyperref}
\setcitestyle{square}
\usepackage{subfigure} 
\usepackage{natbib} 
\usepackage{graphicx,amssymb} 
\usepackage{ulem} 
\usepackage{bm}

\newcommand{\red}{\textcolor{red}}

\newcommand{\ignore}[1]{} 
\newcommand{\Levy}{L\'evy {}}
\def\eq#1{${#1}$} 
\def\EQ#1#2{\begin{equation}{#1}\label{#2}\end{equation}} 
\newcommand{\Chi}{\chi^*} 
\newcommand{\Tau}{\tau^*}
\newcommand{\Intd}{{\rm{d}}} 

\usepackage{gensymb} 

\setcitestyle{square}
\usepackage{subfigure}

\usepackage{enumitem}

\begin{document}

\title{ Large-deviations for spatial diffusion of cold atoms}
\author{Erez Aghion, David A. Kessler, Eli Barkai}
\affiliation{ Department of Physics, Institute of Nanotechnology and Advanced Materials, Bar Ilan University, Ramat-Gan 52900, Israel}

%
%
\begin{abstract}
 Large-deviations theory deals with tails of probability distributions and the rare events of random processes, for example spreading packets of particles.  Mathematically, it concerns the exponential fall-of of the density of thin-tailed systems. Here we investigate the spatial density \eq{P_t(x)} of laser cooled atoms, where at intermediate length scales the shape is fat-tailed. We  focus on the rare events beyond this range, which dominate important statistical properties of the system. Through a novel friction mechanism induced by the laser fields, the  density is explored with the recently proposed nonnormalized infinite-covariant density approach. The small and large fluctuations give rise to a bi-fractal nature of the spreading packet. 
\end{abstract}


\maketitle{}

 In diffusion processes such as Brownian motion, the concentration of particles starting at the origin spreads out like a Gaussian,  which is fully characterized by the mean squared-displacement. This is the result of the widely applicable Gaussian central limit theorem (CLT) \cite{gardiner2009stochastic}. Of no less importance is large-deviations theory \cite{touchette2009large}, which deals with the rare fluctuations of processes such as simple coin tossing random walks  (see e.g., \cite{ellis1999theory}), extreme variations of the surface height in the Kardar-Parisi-Zhang model~\cite{krapivsky2014large} and the tails of the position distribution in single-file diffusion~ \cite{meerson2016large,hegde2014universal}. Mathematically, a prerequisite of the theory is that the cumulant generating function be ``well behaved'', i.e. smooth and differentiable.
 Large-deviations theory works when the decay of the probability of the observable of interest is exponential (see details in \cite{touchette2009large}). However  many systems do not meet this requirement \cite{touchette2009large}, for example L\'evy
 fat-tailed processes  \cite{bouchaud1990anomalous,klafter1996beyond,LevyWalks}, where the decay rate is a power-law. 

This is the case for a cloud of atoms undergoing Sisyphus laser-cooling~ \cite{cohen1990new}, where both theoretically \cite{marksteiner1996anomalous,kessler2012theory} and experimentally  \cite{sagi2012observation}, it was shown that the central part of the spreading particle packet is described by the L\'evy CLT \cite{klafter2011first}. The latter deals with the sum of independent identically-distributed random variables, but unlike the classical Gaussian CLT, here the summands' own distribution is heavy-tailed. As a result, L\'evy's CLT yields an infinite mean squared-displacement for the sum   \cite{klafter2011first}, and consequently also the second cumulant. Large-deviations theory mainly deals with thin-tailed processes where extreme events are rare, but in L\'evy processes these large fluctuations are dominant.  To study the fluctuations in this system, we will show that the relevant tool is the asymptotic moment-generating function, which yields an infinite-covariant density (ICD)  \cite{kessler2012theory,rebenshtok2014non}. We will discuss the generality of this approach and its results below. 

In an experimental situation, diverging moments are unphysical. For example, although the experiment in \cite{sagi2012observation}  shows a nice fit of the particles' density to a symmetric L\'evy distribution,  
 clearly at finite times no particles traveling at finite velocities can ever be found infinitely far from their origin. The finiteness of all the moments requires that the power-law tail of the distribution be cut-off beyond some point. A full characterization of the system demands that this far asymptotic regime be captured correctly, as well as the intermediate asymptotic power-law of the L\'evy CLT. 

\textit{Model.} Sisyphus cooling is controlled by two competing mechanism: the slow decay in time of large momenta due to an anomalous friction force that weakens at large velocities, and random momentum fluctuations which lead to heating \cite{dalibard1989laser,cohen1990new}. Within the framework of the semiclassical approximation, the trajectory of an atom which starts at the origin \eq{x(0)=0}, with \eq{v(0)=0}, is determined by the Langevin equations \cite{marksteiner1996anomalous} (see supplementary material (SM) for a more in-depth review):  \EQ{\dot{v}(t)=\mathcal{F}(v)+\sqrt{2D}\Gamma{(t)},\quad\quad\quad \dot x(t)=v(t),}{eq3:Langevin} where \eq{\mathcal{F}(v)=-v/(1+v^2)} is the deterministic cooling force, in dimensionless units \cite{kessler2010infinite} (physical units in SM). Asymptotically, \eq{\mathcal{F}(v)\sim -v} when \eq{v\ll1} and  \eq{\sim -1/v} when \eq{v\gg1}. \eq{\Gamma(t)} is a Gaussian white-noise with zero mean and \eq{\langle \Gamma(t)\Gamma(t')\rangle=\delta(t-t')}.  \eq{D=cE_R/{U}_0}, where \eq{{U}_0} is the depth of the optical lattice, \eq{E_R} is the recoil energy  and  \eq{c\approx20} is a constant whose precise value is specific to the type of atoms used in the experiment  \cite{cohen1990new,marksteiner1996anomalous}. \eq{U_0}, and hence \eq{D}, may be tuned in the lab, and are the control parameters of the system. Several anomalous statistical predictions of this model, Eq. \eqref{eq3:Langevin}, both in and out of equilibrium, were confirmed in experiments (see e.g., \cite{douglas2006tunable,katori1997anomalous,sagi2012observation}).

We wish to study the large deviations of the probability density function (PDF) of the particles' positions, \eq{P_t(x)} at time \eq{t}. Its Fourier-transform, \eq{\int_{-\infty}^\infty \exp(ikx)P_t(x)\Intd x}, from \eq{x\rightarrow k}, is the moment-generating function  \cite{klafter2011first} \EQ{\hat P_t(k)=1+\sum_{m=1}^{\infty}\frac{(ik)^{2m}}{(2m)!}\langle{x^{2m}}(t)\rangle.}{CharacDef} The strategy we will employ is  to derive the moments of the process, \eq{\langle x^{2m}(t)\rangle}, for \eq{m=1,2,...} (odd moments are zero by symmetry), perform the summation in Eq. \eqref{CharacDef} and invert this function to obtain the density in \eq{x} space.  
Naively, we would expect a normalized density to emerge, but this, as we will show, appears not to be the case.

 \begin{figure}[t]
\centering
\noindent \includegraphics[width=1.0\linewidth]{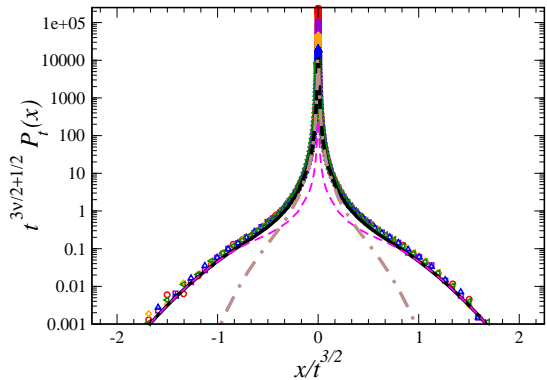}
\caption{{\footnotesize{(Color online) Convergence of the particles position density in Sisyphus cooling, with $D=0.4$ (\eq{\nu=7/6}), to the form of the ICD, Eq. \eqref{eq:7ICDMain}. Langevin simulation results  \cite{Note1} for $t=1000$ are represented by (green, left triangles), $t=1778$ (blue, up triangles), $t=3162$ (orange, diamonds), $t=5623$ (purple, squares) and $t=10000$ (red circles). The scaling limit function \eq{\mathcal{I}(z)}, based on the areal PDFs of the Bessel excursion and the meander, is presented in (solid black line). Asymptotic theory for \eq{(x/t^{3/2})\ll1} (dot-dashed brown line) and \eq{(x/t^{3/2})\gg1} (magenta dashed line), correspond to Eqs. \eqref{ICDApproximationAtSmallZ} and \eqref{ICDApproximationAtLargeZ} respectively.  
Notice how \eq{\mathcal{I}(z)} diverges as \eq{x/t^{3/2}\rightarrow0}, and it is not integrable at this pole. }}}
\label{fig1:ICD}
\end{figure}


\textit{Scaling arguments for a nonnormalizable state.} An initial insight into the position distribution, \eq{P_t(x)}, may be gained as follows: Let \eq{W_t(x,v)} be the phase space distribution of the diffusive particle packet, at time \eq{t}.  Since for large \eq{v} the friction vanishes from Eq. \eqref{eq3:Langevin}, in this case we expect a scaling \eq{v \propto t^{1/2}}. By  integration over time, this implies \eq{x \propto t^{3/2}}. Based on these scaling arguments we may write 
\eq{W_t(x,v) \sim t^{\xi} f\left(x/t^{3/2},v/t^{1/2}\right)}.
To determine the exponent \eq{\xi} we may use a simple argument (though it can be derived also rigorously): We note that when \eq{D<1} the marginal velocity equilibrium density is \cite{douglas2006tunable,lutz2004power,kessler2010infinite,holz2015infinite,dechant2016heavy}  \EQ{\lim_{t\rightarrow\infty}\mathbf{P}_t(v)\rightarrow \mathbf{P}_{eq}(v)\sim |v|^{-1/D},\quad(\mbox{when}\quad v\gg1).}{PEqVelocity} The range  \eq{D>1} is a heating phase, where an equilibrium state does not exist, hence we leave it out of the context of this work. By definition, this velocity density is related to the phase space distribution via \EQ{\mathbf{P}_{eq}(v)=\lim_{t\rightarrow\infty} t^{\zeta+3/2} \int_{-\infty}^{\infty} f\left(\frac{x}{t^{3/2}},\frac{v}{t^{1/2}}\right)\Intd \left(\frac{x}{t^{3/2}}\right),}{ScalingAnsatz} hence from Eqs. (\ref{PEqVelocity},\ref{ScalingAnsatz}) we find \eq{\zeta=-3/2-1/(2D)}. 

Using this result, integration of the scaling solution over 
velocity  yields $P_t(x) \sim \mathcal{I}(z)/t^{1+1/(2D)}$ where~$z=x/t^{3/2}$. This suggests, and indeed our rigorous theory shows, that there exists a limit such that   \EQ{\mathcal{I}(z)=\lim_{t\rightarrow\infty}t^{1+1/(2D)}P_t(x).}{eq:1ICDDef}  
This limit is interesting since if we integrate Eq. \eqref{eq:1ICDDef} over $dz=dx /t^{3/2}$ we get from the normalization of $P_t(x)$, that the integral $\int_{-\infty} ^{\infty} \mathcal{I}(z) \Intd z\rightarrow \infty$. It follows that $\mathcal{I}(z)$ is not a normalized density, but rather a scaling solution that captures the non-unifrom convergence of the packet of particles. As we discuss below this scaling limit is not unique, but luckily there exists only one more scaling limit to the problem, and that is described by the well known L\'evy CLT. In that sense the nonnormalised state \eq{\mathcal{I}(z)}, being a limiting solution, is complementary to the CLT.  
In Fig. \ref{fig1:ICD}, we present simulation data from the cold atoms system with \eq{D=0.4} \footnote{\label{Simulations} Simulations were performed   using standard Euler-Mayurama integration \cite{gardiner2009stochastic} of Eq.~\eqref{eq3:Langevin} with a step size of $\Delta t=0.01$, for $10^5$ particles.}, which shows nice convergence with increasing time to the theory.  


%
\textit{Excursions to untangle Langevin dynamics}. 
The derivation of our main results uses a connection between the properties of constrained stochastic paths and Langevin dynamics, established in \cite{marksteiner1996anomalous,barkai2014area}. Let the times \eq{t_1,t_2,...t_n} denote the zero crossings of the stochastic process \eq{v(t)}, Eq. \eqref{eq3:Langevin}. The time intervals between the crossing events, \eq{\tau_1=t_1-0, ... \tau_n= t_{n} - t_{n-1}}, are independent identically-distributed random variables, a property which is due to the Markovian Langevin process under investigation. The total measurement time is  \eq{t=\sum_{i=1}^{n}\tau_i+\tau^*}, where \eq{\Tau} is the duration of the last interval, in which the velocity does not return to zero. The displacement accumulated by the particle during each interval is~\eq{\chi_i=\int_{t_{i-1}}^{t_{i-1}+\tau_i} v(t)\Intd t} (for the last step, \eq{\Chi=\int_{t-\Tau}^{t} v(t)\Intd t}), and the final random position of the particle at time \eq{t} is given by the sum \eq{x(t)=\sum_{i=1}^{n}{\chi_i}+\chi^*}.    
Note that in this construction, the velocity path in all but the last interval starts and ends at zero, and is strictly positive or negative in between, hence the \eq{\tau_i}s are determined by the first-passage time (to the velocity origin) distribution: \eq{g(\tau)\approx g^*\tau^{-3/2-1/(2D)},} for large \eq{\tau}  \cite{kessler2010infinite,barkai2014area}. The slow decaying power-law tail of this function, means that the duration of the last step might be as long as the sum of all the prior ones and it cannot be neglected. 
This is clearly a consequence of the weak friction at large velocities, that allows for very long flights without velocity zero crossings. 

\begin{figure}[t]
  \includegraphics[width=1\linewidth,height=1.92cm]{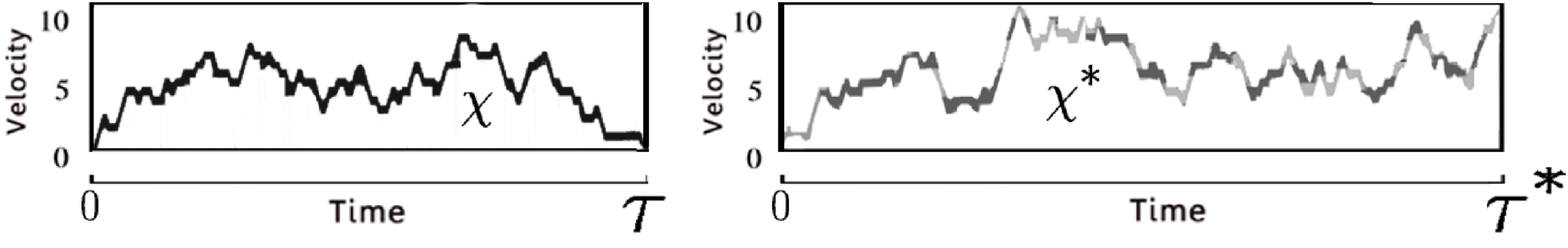}
  \caption{{\footnotesize{\textbf{On the left:} A Bessel excursion in velocity space, follows \eq{\dot{v}=-1/v+\sqrt{2D}\Gamma(t)}, with the path constrained to start at \eq{v(0)=\epsilon} and end at \eq{v(\tau)=0}, and remain strictly positive in the time interval \eq{(0,\tau)}. The random area under the path is \eq{\chi}. \textbf{On the right:} A velocity Bessel meander, with duration \eq{\Tau} and area \eq{\Chi}, starts at \eq{v(0)=\epsilon} and remains positive, while the final value  \eq{v(\Tau)} is random.}}}
\label{SchematicDescription} 
\end{figure}

Each segment of the path \eq{v(t)}, prior to the last, (i.e. between zero crossings), is approximated by a Bessel excursion in velocity space \cite{barkai2014area,kessler2014distribution}  (see Fig. \ref{SchematicDescription}). 
An excursion in the time interval \eq{[0,\tau_i]}, is a stochastic trajectory which is constrained to begin close to the velocity origin, at \eq{v(0)=\epsilon\rightarrow0}, end at \eq{v(\tau_i)=0}, and never reach zero between \eq{(0,\tau_i)} (see e.g., \cite{louchard1984kac,pitman1999brownian,majumdar2015effective}). \eq{\chi_i\propto \tau_i^{3/2}}, is the area under the \eq{i}'th excursion, which is naturally correlated to its duration, since longer duration means larger displacement. The last segment, where the velocity path is not conditioned at its final point, is called a velocity Bessel meander \cite{durrett1977weak,barkai2014area} (Fig. \ref{SchematicDescription}). 
The term Bessel derives from the fact that for \eq{v\gg1}, Eq.~\eqref{eq3:Langevin} is mathematically related to the Bessel process which describes the radial component of Brownian motion in arbitrary dimensions \cite{schehr2010extreme,martin2011first,kessler2012theory,font2016percolation}. Clearly the statistics of $\chi$ and the zero crossing times, $\tau$, determines the random position of the particle, $x(t)$. Since the \eq{\tau}s are independent and identically distributed, the zero crossings form a renewal process  \cite{barkai2014area,montroll1965random}, which allows us to analyze the problem analytically. 

 In the SM, we find the following asymptotic expression for the \eq{2m}'th moment of the particles' positions, valid for \eq{m\geq1} at long-times, in the range \eq{1/5<D<1} (the range \eq{D<1/5} is addressed below, details on the prefactor \eq{g^*/\langle\tau\rangle} are provided in the SM):  
\small  	
\begin{align} 
&\qquad\qquad\qquad\qquad\left\langle x^{{2m}}(t)\right\rangle\approx\frac{{g^*}}{\left\langle\tau\right\rangle}t^{3m-3\nu/2+1}\times\nonumber\\ 
&\left[\frac{\langle\chi^{{2m}}\rangle_E}{|(3m-3\nu/2)(3m-3\nu/2+1)|}+\frac{2\langle\chi^{{2m}}\rangle_M}{3\nu|(3m-3\nu/2+1)|}\right],  
\label{eq25}  
\end{align}
\normalsize 
where \EQ{\nu=\frac{1}{3D}+\frac{1}{3},\qquad ({2/3<\nu<2}).}{eq:8v32NuDef} We denote by  \eq{\langle\chi^{{2m}}\rangle_E=\int_{-\infty}^\infty \chi^{{2m}}B_E(\chi)\Intd \chi} the \eq{2m}th moment of the areal distribution of the Bessel excursion, \eq{B_E(\chi)}, in the time interval  \eq{[0,1]}. Similarly, we denote by \eq{B_M(\chi)} and \eq{\langle\chi^{{2m}}\rangle_M} the distribution and moment, respectively, of the meander in same time interval. Note that, importantly, Eq. \eqref{eq25} does not apply for \eq{m=0}. The exact scaling of the moments in Eq. \eqref{eq25} immediately suggests that the particle density may converge, after rescaling, to the limit function suggested by Eq. \eqref{eq:1ICDDef}; we provide general arguments in the end of the paper.  
 
\textit{Nonormalizable limit function for the PDF}. Using the long-time asymptotic moments provided in Eq. \eqref{eq25}, in the moment-generating function, Eq. \eqref{CharacDef}, yields an approximation	for \eq{\hat{P}_t(k)}, which we denote \eq{\hat{P}_t^A(k)}, valid at long times: 
\small
\begin{align}
&\qquad\hat P_{t}^A(k)=1+t^{-3\nu/2+1}\sum_{n=1}^{\infty}\frac{(-1)^n{g^*}(kt^{3/2})^{2n}}{\langle\tau\rangle(2n)!}\times\nonumber\\ 
&\left[\int_{-\infty}^\infty  \chi^{2n}B_E(\chi)\Intd \chi\left(\frac{1}{{3n}-3\nu/2}-\frac{1}{3n-3\nu/2+1}\right)+\right.\nonumber\\
&\qquad\left.\int_{-\infty}^\infty  \chi^{2n}B_M(\chi)\Intd \chi\frac{2}{3\nu(3n-3\nu/2+1)}\right].
\label{Gen1} 
\end{align}
\normalsize
Rearranging, and using the Taylor expansion \eq{\cos\left(\omega^{3/2} y\right)=\sum_{n=0}^\infty(-1)^n\left(\omega^{3/2} y\right)^{2n}/(2n)!} for the summation, we obtain  
\small
\begin{align}
\hat P_{t}^A(k)&=1+\frac{{g^*}t^{-3\nu/2+1}}{\langle\tau\rangle}\int_{-\infty}^\infty\Intd \chi\int_0^1\Intd{}\omega\left[\cos\left(\omega^{3/2} k{\chi} t^{3/2}\right)-1\right]\nonumber\\ 
&\times\left[\frac{B_E({\chi})}{\omega^{3\nu/2-1}}+\frac{2B_M({\chi})-3\nu B_E({\chi})}{3\nu\omega^{3\nu/2}}\right].  
\label{Gen125} 
\end{align}
\normalsize
Immediately below, taking the inverse-Fourier transform from \eq{k\rightarrow x}, we drop  
the term proportional to \eq{\delta(x)}, since this analysis applies only at large \eq{x}. Calculating the integral over \eq{\omega}, we now obtain the limit function \eq{\mathcal{I}(z)}, explicitly, which describes the particle packet from its relation to the appropriately rescaled density via Eq.~\eqref{eq:1ICDDef}. In the limit   \eq{z=x/t^{3/2}\ll1}:  
\EQ{\mathcal{I}({z})\approx\frac{{g^*}}{3\langle\tau\rangle}\langle{|\chi|^{\nu}\rangle}_E{|{z}|^{-\nu-1}},}{ICDApproximationAtSmallZ}
where \eq{\langle|\chi|^\nu\rangle} is the \eq{\nu}'th absolute-moment of the excursion \cite{kessler2014distribution}. Note that this equation means that \eq{\mathcal{I}(z)} is nonintegrable around the origin. 
For \eq{z\gg1}: 
\begin{equation} 
\centering
{\mathcal{I}({z})\approx\frac{4{g^*}}{9\nu\langle\tau\rangle}{{|z|}^{-\nu-1/3}}\int_z^\infty {|\chi|^{\nu-2/3}}B_M(\chi)\Intd\chi.}
\label{ICDApproximationAtLargeZ}
\end{equation} 

The function \eq{\mathcal{I}(z)} is called the infinite-covariant density (ICD), of the spatial diffusion of the cold atoms. The term \textit{infinite}, means that it is nonnormalizable, despite being a limit function of the (obviously normalized) PDF. The term \textit{covariant} refers to the fact that it is a function of the scaled variable \eq{x/t^{3/2}}.  We were able to obtain this nonnormalizable solution from the standard moment-generating function since in Eq. (\ref{CharacDef}) we summed over the long-times asymptotic approximation, rather then the exact moments. Eq. \eqref{ICDApproximationAtLargeZ} is the long-time asymptotics of the tail of the PDF, and in that sense it describes the rare fluctuations of the system.  Fig. \ref{fig1:ICD}, confirms the convergence of Langevin simulation results, obtained by numerical integration of Eq. \eqref{eq3:Langevin}  at increasing times, to the nonnormalised density and its asymptotic approximations, Eqs. (\ref{ICDApproximationAtSmallZ},\ref{ICDApproximationAtLargeZ}). These asymptotic limits are controlled exclusively by the excursions for $z\ll1$ and the meander for $z\gg1$, thus the far tail is 
described by a path that did not switch its velocity direction for a duration of the order of measurement time. Clearly, this is a rare event.

For every \eq{z}: 
\begin{widetext} 
\begin{equation} 
\mathcal{I}(z)=\frac{2{g^*}}{3\langle\tau\rangle}\frac{1}{|{z}|^{\nu+1}}\left[\int_{|{z}|}^\infty B_E(\chi)|\chi|^{\nu}\Intd \chi+|z|^{2/3}\int_{|{z}|}^\infty \left(\frac{2}{3\nu}B_M(\chi)-B_E(\chi)\right){|\chi|^{\nu-2/3}}\Intd \chi\right].  
\label{eq:7ICDMain} 
\end{equation}
\end{widetext}
Explicit expressions for the areal distributions, \eq{B_E(\chi)} and \eq{B_M(\chi)}, used with Eq. \eqref{eq:7ICDMain} to plot the theory in Fig. \ref{fig1:ICD}, are provided in the SM. The ICD, \eq{\mathcal{I}(z)}, gives the long times limit of all the absolute integer and fractional moments \eq{\langle |x|^q\rangle} of order \eq{q>\nu}. Remarkably, this  also includes the second moment, generally considered 
in many experiments as the typical characterization of a diffusion process. Looking back at Eq. \eqref{eq25}: The mean squared-displacement, which is sensitive to the large fluctuations and the tails of the PDF, is obtained via \eq{\langle x^2(t)\rangle=t^{4-3\nu/2}\langle z^2\rangle_{\mathcal{I}}}, where \eq{\langle	 z^2\rangle_{\mathcal{I}}=\int_{-\infty}^{\infty} z^2\mathcal{I}\left(z\right)\Intd z}. 
For every \eq{q\in\mathbb{R}}: if \eq{\mathcal{I}(z)}, Eq. \eqref{eq:7ICDMain}, is integrable with respect to \eq{|z|^q}, then the ICD  determines \eq{\langle |x|^q(t)\rangle} (i.e., the \eq{q}'th absolute-moment \footnote{Eq.~\eqref{eq25} is analytically continued to absolute odd and fractional moments \eq{\langle|x|^q\rangle}, for \eq{\nu<q\in\mathbb{R}}, by replacing \eq{2m\rightarrow q} and using \eq{\langle|\chi|^{{q}}\rangle_E} and \eq{\langle|\chi|^{{q}}\rangle_M}.}) via \eq{\langle |x|^q(t)\rangle=t^{3q/2-3\nu/2+1}\langle |z|^q\rangle_{\mathcal{I}}}, where \eq{\langle |z|^q\rangle_{\mathcal{I}}=\int_{-\infty}^{\infty} |z|^q\mathcal{I}\left(z\right)\Intd z}. Contrarily, when  \eq{q\leq\nu},  \eq{\mathcal{I}(z)} is nonintegrable with respect to the observable \eq{|z|^q}, hence the moments which are less sensitive to large fluctuations are given by the L\'evy distribution, as was found in \cite{kessler2012theory}. This second long-time limit function has the scaling shape   \eq{t^{-1/\nu}\mathcal{L}_\nu (x/t^{1/\nu})} \cite{kessler2012theory}. 
For all the absolute-moments we find the bi-scaling behavior 
\begin{equation} 
 \langle |x|^q(t)\rangle\propto \begin{cases} t^{q/\nu}\quad\quad\quad\quad\quad\quad  q<\nu\\ t^{3q/2-3\nu/2+1}\quad\quad q>\nu\end{cases}. 
\label{Scaling}
\end{equation} 
Such multifractality is known as strong anomalous diffusion~ \cite{castiglione1999strong\iffalse,andersen2000simple\fi}. It represents the multi-scaling nature of the underlying PDF. Note that as \eq{q\rightarrow\nu} from above, the coefficient of \eq{\langle|x|^q\rangle}, given by the analytic continuation of  Eq. \eqref{eq25}, diverges. The same happens when evaluating the moments using the L\'evy scaling function, and approaching \eq{\nu} from below. 

The derivation of \eq{\mathcal{I}(z)}, Eq. \eqref{eq:7ICDMain}, was performed in the limited range of \eq{D} where the variance is provided by the ICD. However, the scaling arguments at the beginning of this letter suggest that such a function should be found whenever the power-law equilibrium state in velocity space, Eq. \eqref{PEqVelocity}, exists, namely for all \eq{0<D<1}. Indeed, one can show that the ICD is valid also in the range \eq{0<D<1/5}, where one finds that \eq{\langle x^2\rangle} grows linearly in time and the central part of the spreading packet is Gaussian. Even in this Gaussian regime, standard large-deviations theory does not apply and instead, the ICD given by Eqs. (\ref{eq:1ICDDef},\ref{eq:7ICDMain}) insures the finiteness of large moments, beyond the mean squared-displacement. There is a delicate matching problem between the Gaussian packet and the pole of the ICD that describes the rare events,  which we will address elsewhere. 
%
 %

\textit{Generality of the infinite-covariant density approach.} We suggest that ICDs may be naturally related to multi-fractality (see physical examples below). In particular we now derive a rather general relation between exponents describing the 
bi-fractal moments, the central part of the packet (i.e., the bulk fluctuations,  described by the L\'evy CLT), and the exponents describing
the ICD. When absolute-moments of order \eq{q>q_c}, where \eq{q_c>0} defines some critical moment, scale faster in time than smaller ones, a scaling function  \eq{\mathcal{I}(\tilde{z}=x/t^\alpha)} describes the large fluctuations at long times via~\eq{\langle |x|^{q}(t)\rangle\rightarrow t^{q\alpha - \beta+\alpha}\int_{-\infty}^\infty |\tilde{z}|^{q}\mathcal{I}(\tilde{z})\Intd \tilde{z}} (\eq{\beta>\alpha>0}). In this  case one may find that~\eq{\mathcal{I}(\tilde{z})=\lim_{t\rightarrow\infty}t^\beta P_t(x)}, where \eq{P_t(x)} is the normalized PDF. This limit function is hence a nonnormalizable ICD (since  obviously \eq{\langle x^{0}(t)\rangle =1}, then~\eq{ \int_{-\infty}^\infty \mathcal{I}(\tilde{z})\Intd{\tilde{z}}\rightarrow\infty}). In the case that around the origin the PDF is represented by a \Levy distribution of the form \eq{t^{-1/\gamma}\mathcal{L_\gamma}(x/t^{1/\gamma})} \cite{klafter1994levy}, one will find (by ``stitching'' this limit function and the ICD at a central region of \eq{x}, as in   \cite{rebenshtok2014non}) the following relation between the scaling exponents: \eq{\alpha-\beta+\alpha\gamma=1.} Indeed, in our case, \eq{(\alpha,\beta)=(3/2,1+1/(2D))} gives the correct \eq{\gamma=(1+D)/3D=\nu}. In~ \cite{dentz2015scaling}, for example, the authors study a nonlinearly coupled continuous-time random walk with 
\eq{(\alpha,\beta)=(\bm{\alpha},\bm{\alpha}+\bm{\beta}-1)}, which according to our analysis yields \eq{\gamma=\bm{\beta}/\bm{\alpha}}  (\eq{\bm{\alpha},\bm{\beta}} refer to the parameters in this Ref.). Our prediction  is consistent with the result of their analysis.
  A more general relation links the exponents \eq{\alpha,\beta},  of the ICD and the central power-law where \eq{x\sim t^{1/\gamma}}, to the critical moment of the bi-scaling, \eq{q_c}: \eq{\alpha-\beta+q_c\alpha=q_c/\gamma}. This is consistent {e.g.}, with the exponents found for transport on \eq{2}-dimensional \Levy quasicrystals, studied in \cite{buonsante2011transport}. The agreement with \cite{dentz2015scaling,buonsante2011transport} suggests an ICD in these systems too.  

While non-analytical behavior of the moments  raises a red flag for standard large-deviations theory, it promotes the use of the ICD approach.  Finding this function is crucial for characterizing the rare events. 
The limit law given by the ICD in Eq.  \eqref{eq:1ICDDef} for the rescaled PDF   provides an alternative to the large-deviations principle, according to which the decay of the tails in thin-tailed systems may be  controlled by some rate function \eq{Q(x/t)}, such that   \eq{Q(x/t)=\lim_{t\rightarrow\infty}\ln\left[P_t(x)\right]/t}. 

{\textit{Discussion.}} 
 CLTs play an important role in statistical physics, but of no less importance may be the proper characterization of the  deviations from them. 
The ICD was previously found, for example, for different models of \Levy walks \cite{rebenshtok2014non,rebenshtok2014infinite}. Since dual scaling of the moments and fat tailed distributions are very common, we speculate that ICDs will describe a large class of systems, e.g., L\'evy glasses   \cite{bernabo2014anomalous},  fluctuating surfaces \cite{zamorategui2016distribution}, motion of tracer particles in the cell \cite{gal2010experimental} and diffusion on
lipid bilayers \cite{\iffalse burioni2012scattering,\fi krapf2016strange}.  To identify the ICDs in these diverse systems requires further work. Here, we have derived the ICD from the semiclassical description of cold atoms. This system is unique since it allows us, by tuning the intensity of the lasers, to find regimes where large deviations in the tails are non-negligible. In this case the rare events are important since 
they determine prominent statistical properties of the system, such as the mean squared-displacement. Our ICD is complementary to L\'evy's CLT in the sense that it solves the serious problem of the diverging variance expected by the L\'evy distribution, although the latter insures the normalizability of the PDF. A full description of the system requires both functions. 

Our work leaves open many interesting  questions. One is the shape of the ICD when prior to measurement, the spreading particles are left to relax by interacting with the lasers in a spatial trap for some time \eq{t^*}, where \eq{t^*\gg t}. Our results apply in the opposite limit. In a previous work, Dechant and Lutz \cite{dechant2012anomalous} find not bi-scaling, but tri-scaling of the moments in this case.  In general, the ICD may depend on the protocol of the preparation of the system. In particular, the dependence on $t^*$ leads to aging effects, i.e., transport that depends on the preparation time. 
Finally, we point out that the function, \eq{f(z,\tilde{v})} (where \eq{\tilde{v}=v/t^{1/2}}) in Eq. \eqref{ScalingAnsatz}, is itself an ICD, as it is clearly not normalizable. Elucidating the properties of this ICD is an important future goal. 

This work was supported by the Israel Science Foundation. 

\bibliographystyle{apsrev4-1}
\bibliography{./bibliography} 

\newpage
\onecolumngrid
\section
{Supplementary Material for: \\ Large deviations for spatial diffusion of cold atoms}

\subsection{A. Sisyphus cooling} 
\label{DerivationMechanism} 

Sisyphus cooling \cite{cohen1990new} uses two coherent orthogonal, linearly-polarized laser beams, in a $1$-dimensional lin$\perp$lin configuration. The counter propagating lasers are projected onto a packet of hydrogen-like atoms (e.g., \eq{{}^{87}}Rb), creating an optical lattice. The cooling mechanism is driven by the coupled effect of periodic potential energy shifts, experienced by the particle as it moves along the lattice, and precisely timed, repeated, optical pumping events, which make the atom effectively  move constantly ``up'' a potential hill (and hence the name Sisyphus cooling is appropriate). This induces a secular loss of kinetic energy for the atoms. In the semicalssical approximation one performs an average over the spatial modulation of the optical lattice, which works especially well in the limit of relatively fast particles.

In physical units, the deterministic damping force induced by the lasers may be written, as \cite{cohen1990new,marksteiner1996anomalous}:  
\EQ{F(\tilde{p})=-\frac{{\bar{\alpha}\tilde{p}}}{1+(\tilde{p}/p_c)^2},\qquad\mbox{where}\qquad\bar{\alpha}=\frac{12\pi^2\hbar|\delta|}{\lambda^2m\Gamma},\qquad\mbox{and}\qquad p_c=m\lambda/(4\pi\tau_p).}{CMForce} 
Here, \eq{\tilde{p}} is the momentum of the atom, and $p_c$ is set by the velocity for which the atom travels the distance between two maximum points of the optical lattice in the time span of  one optical pumping; \eq{\tau_p={9}/(2\Gamma s_0)}. The spatial periodicity of  the optical lattice is half the wavelength, \eq{\lambda}, of the lasers  \cite{cohen1990new}. The dimensionless saturation parameter; $s_0={2\Omega_R^2}/({4\delta^2+\Gamma^2})$,  is dependent on the parameters of the laser and the lifetime of the excited state of the atom. The Rabi frequency is $\Omega_R$ and \eq{\delta=\omega_l-\omega_R} is the detuning  between the laser frequency and the atom's electronic transition frequency. 
		
Since Sisyphus cooling is driven by quantum effects, instead of a monotonic decrease of the particle's velocity, one finds momentum fluctuations, which in the semiclassical approximation are treated as a Gaussian white-noise in \eq{p} space  \cite{cohen1990new}. The time-development of the phase-space density, \eq{W_{\tilde{t}}({\tilde{x}},{\tilde{p}})}, at time \eq{{\tilde{t}}} is given by  Kramer's Eq.  \cite{marksteiner1996anomalous}, \EQ{\frac{\partial W_{\tilde{t}}({\tilde{x}},{\tilde{p}})}{\partial {\tilde{t}}}+{\tilde{p}}\frac{\partial W_{\tilde{t}}({\tilde{x}},{\tilde{p}})}{\partial {\tilde{x}}}=\left[\tilde{D}\frac{\partial^2}{\partial^2 {\tilde{p}}}-\frac{\partial}{\partial {\tilde{p}}}F({\tilde{p}})\right] W_{\tilde{t}}({\tilde{x}},{\tilde{p}).}}{eq:5Kramers} The amplitude of the momentum fluctuations has two components:
\EQ{\tilde{D}=D_1+\frac{D_2}{1+({\tilde{p}}/p_c)^2},\qquad \mbox{where}\qquad D_1=11mE_R/(2\tau_p)\qquad \mbox{and} \qquad {D_2=9\pi^2U_0^2/(\lambda^2s_0\Gamma)} 
.}{CM7} Here, \eq{E_R=2\pi^2h^2/(m\lambda^2)} is the recoil energy, and \eq{U_0=\frac{2}{3}\hbar\delta s_0} is the depth of the optical lattice \cite{cohen1990new}. The first component of the fluctuations, $D_1$, is the result of the recoil due to the emission of the photon during the optical pumping. The second component relates to emissions occurring in ``the wrong points'' on the optical lattice, which result in
temporary gains of kinetic energy (when the atom ``slides'' down the potential). For slow particles, \eq{\tilde{D}\sim  D_1 + D_2}, while for fast particles, \eq{\tilde{D}\sim  D_1}. From Eq. (\ref{CMForce}),  the cooling-force is small when acting on
fast particles, hence these particles tend to remain fast for long times and in the range of D in which we are interested in 
the main text, they dominate the statistical properties of the diffusing packet.
We therefore neglect the contribution of  $D_2$, and use   $\tilde{D}= D_1$ (this agrees with simulations, see e.g.  \cite{dechant2016heavy}). 

Finally, we work in dimensionless units, \eq{(x,p,t)}, where  \cite{barkai2014area,dechant2016heavy}: 
\EQ{p=\tilde{p}/p_c, \qquad{t= \tilde{t}\bar{\alpha}}\qquad \textrm{and} \qquad {x= \tilde{x}\bar{\alpha}m/p_c}. }{DimlessUnits}
Note that we take the particle's mass to be $m=1$ for convenience, hence in the main text \eq{p=v}, where \eq{v} is the dimensionless velocity (and we can set \eq{p_c=v_c=1}). In these units,  \eq{D=\tilde{D}/\left[p_c^2\bar{\alpha}\right]=cE_R/U_0}, where \eq{c\approx22}, and the  dimensionless Langevin Eq. (\eq{1}), in the main text, is the equivalent of the Kramers equation \eqref{eq:5Kramers} above. Note that the constant \eq{c} may differ between different theoretical works and experiments, since the exact numbers in Eq. \eqref{CM7} depend on the details of the estimation of the noise in different experimental setups, and the particular  atomic transition. In the main text we consider the range $0 < D <1$, which translates to \eq{0<D_1/[p_c^2\bar{\alpha}]<1.}

\subsection{B. Excursions approach for solving the Langevin equation}
\label{ModifiedMWEqautionDerivation}
Here we present the derivation of the relation between the probability density functions (PDFs) of the area under the Bessel excursion and Bessel meander, and the Sisyphus-cooled particles' position PDF, \eq{P_t(x)}, at time \eq{t}  \cite{barkai2014area}. We call this relation the modified Montroll-Weiss equation (see \cite{montroll1965random}). We  repeat this derivation here, which we first presented in \cite{barkai2014area}, since it is a bit different then the famous original relation (see [SM1] for a  review),  due to the specific treatment given to the meander (which is separate from the excursions).  This modified equation is the starting point for our calculation of the integer moments  \eq{\langle|x|^{2m}\rangle} for \eq{m\geq1}, Eq. (\eq{6}) in the main text. As mentioned there, we use the zero crossings of the Markovian process, \eq{v(t)} (Eq. (\eq{1}) in the main text),  to define the waiting times
\eq{\left\{\tau_i\right\}} and the corresponding excursions  with areas \eq{\left\{\chi_i\right\}}.  The joint distributions for the area and the duration of an excursion is  
\begin{equation}
\Phi_E\left(\chi,\tau\right)\sim g(\tau)\phi_E(\chi|\tau).   
\label{AppModMW:DeltaM}
\end{equation}
Here \eq{g(\tau}) is the first-passage time PDF of the process \eq{v(t)}, from \eq{\epsilon} to zero (eventually \eq{\epsilon} is taken to zero and cancels out, see \cite{barkai2014area}), and $\phi_E(\chi|\tau)$ is the conditional PDF for \eq{\chi}  given \eq{\tau}. 
This density has the scaling form \cite{barkai2014area} 
\begin{equation} 
\phi_E\left(\chi|\tau\right)\sim\frac{1}{\tau^{3/2}}B_E\left(\frac{\chi}{\tau^{3/2}}\right). 
\label{APPMeander}
\end{equation}  

Let $\eta_s(x,t) {\rm d} t {\rm d} x$ be  the probability that the particle crossed the zero velocity state, $v=0$, for the $s$th time in the time interval $(t,t+{\rm d} t)$, and that its position is in the interval $(x,x + {\rm d} x)$.
This probability is related to the the previous crossing via 
\begin{equation}
\eta_s (x,t) = \int_{-\infty}^\infty  {\rm d} { \chi} \int_0 ^t {\rm d} \tau\, 
\eta_{s-1} \left( x - {\chi} , t - \tau \right) {1 \over   \tau^{{3/2}}} B_E \left( {\chi \over  \tau^{{3/2}} }\right) g \left( \tau \right),
\label{eq08}
\end{equation} 
where we have used  Eqs. (\ref{AppModMW:DeltaM},\ref{APPMeander}). Changing variables from  ${\chi}\rightarrow \zeta \tau^{{3/2}}$
we obtain 
\begin{equation}
\eta_s(x,t) = \int_{-\infty} ^\infty\!\! {\rm d} \zeta \int_0 ^\infty\!\! {\rm d} \tau\, \eta_{s-1} \left( x - \zeta \tau^{3/2}  , t- \tau\right)B_E\left(\zeta \right)g\left( \tau \right) . 
\label{eqMW09}
\end{equation}
The process is now described by a sequence of waiting times $\tau_1, \tau_2,...$ and the corresponding scaled displacements  
$\zeta_1, \zeta_2 , ...$. The displacement in the $s$th interval is  
\begin{equation}
{\chi}_s = \zeta_s\tau_s^{3/2}
\label{eqMW09a}
\end{equation}
The advantage of this representation of the problem, in terms of the pair of microscopic stochastic variables 
$\tau,\zeta$ (instead of the correlated pair $\tau,\chi$), is that we may treat $\zeta$ and $\tau$ as independent random variables whose corresponding PDFs are $g(\tau)$ and $B_E(\zeta)$ respectively. Here  $\tau>0$ and $-\infty<\zeta<\infty$. The initial condition $x=0$ at time $t=0$ implies $\eta_0(x,t) = \delta(x) \delta(t)$.
The probability, $P_t(x)$, of finding  the particle in $(x,x+ {\rm d} x)$ at time $t$, is obtained from the relation 
\begin{equation} 
P_t(x) =
\sum_{s=0} ^\infty \int_{-\infty} ^\infty  {\rm d} \zeta \int_0 ^t {\rm d} \tau^* \, \eta_s \left( x - \zeta {\tau^*}^{3/2} , t - \tau^* \right)B_M\left(\zeta \right) w\left( \tau^* \right).
\label{eqMW10}
\end{equation}
Here we used Eq. \eqref{APPMeander}, and since the last jump event took place at $t - \tau^*$, and in the time period $(t-\tau^*,t)$ the particle did not cross the velocity origin, as mentioned, the last time interval in the sequence is described by a meander. By definition;  $w(\tau^*)= 1-\int_0 ^\tau g(\tau^*) {\rm d} \tau^*$ is the survival probability. The probability to have a meander with an area \eq{\Chi} underneath it, and  duration $\Tau$, is   
\EQ{\Psi_M\left(\chi^*,\tau^*\right)=\frac{1}{{\Tau}^{3/2}}w(\tau^*)B_M(\frac{\chi^*}{{\tau^*}^{3/2}}).}{AppModMW:DeltaMeanderM} 
We provide explicit expressions for  \eq{B_E(\zeta)} and \eq{B_M(\zeta)} below.   
The summation in Eq. (\ref{eqMW10}) is performed over all the possible realizations with $s$ returns to the velocity origin, $v=0$.  
In Laplace $t \to u$ and Fourier $x \to k$ spaces, using the convolution theorem and Eq. (\ref{eqMW09}), we find 
\begin{equation}
\hat{\eta}_s\left( k , u \right) =\hat{ \eta}_{s-1} \left( k , u \right) LT\left[ \hat{B}_E \left( k \tau^{3/2} \right) g \left( \tau \right) \right], 
\label{eqMW11}
\end{equation}
where $LT[\cdot]$ means Laplace transform, and 
\begin{equation}
\hat{B}_E \left( k\tau^{3/2} \right) = \int_{-\infty} ^\infty \exp\left( i k \zeta{\tau^{3/2}}  \right) B_E (\zeta) {\rm d} \zeta.
\label{eqMW12}
\end{equation}
Hence 
\eq{
\hat{\eta}_s (k,u) = \hat{\Phi}_E(k,u) \hat{\eta}_{s-1} (k,u) .} 
This implies that
\begin{equation} 
\hat{\eta}_s(k,u) = \left[ \hat{\Phi}_E(k,u)\right]^s,
\label{eqMW13a}
\end{equation} 
reflecting the renewal property of the underlying random walk. 
Summing the Fourier and Laplace transform of Eq. (\ref{eqMW10}), applying the 
convolution theorem and using Eq.
(\ref{eqMW13a}),  we find the modified Montroll-Weiss equation for the Fourier and Laplace transform of $P(x,t)$:  
\begin{equation}
{\hat{P}_u(k) = {\hat{\Psi}_M (k,u)   \over 1 - \hat{\Phi}_E\left({k,u} \right)}.}
\label{eqMW14}
\end{equation} 
%

\subsection{C. Area distribution under the Bessel excursion and Bessel meander} 
\label{Bessel}

\begin{figure} 
\centering
\includegraphics[scale=0.92]{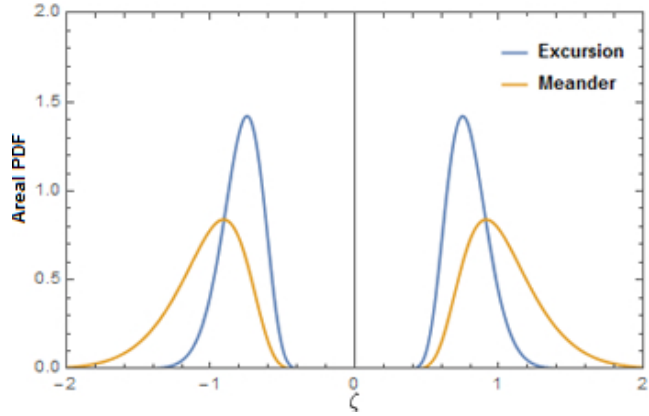}
\caption{{\footnotesize{(Color online) The areal distributions, \eq{B_E(\chi)} and \eq{B_M(\chi)}, of the area under the (positive and negative) Bessel excursion and Bessel meander with \eq{D=0.4}, are presented in (blue) and (orange), respectively.}}}
\label{fig0}
\end{figure} 

The symmetric area distribution \eq{B_E(\zeta)}, for the scaled area \eq{\zeta=\chi/\tau^{3/2}} under the Bessel excursion of duration \eq{\tau} (which takes into account both the paths that always remain positive, and those which remain negative) is \cite{barkai2014area,kessler2014distribution} 
\begin{eqnarray} 
B_E(\zeta) &=& -\frac{\Gamma(1+\alpha)}{4\pi |\zeta|}\left(\frac{4D^{1/3}}{|\zeta|^{2/3}}\right)^{\frac{3\nu}{2}+1}\sum_{k} [{d}_k]^{2}\left[\Gamma\left(\frac{5}{3}+\nu\right)\sin\left(\pi\frac{2+3\nu}{3}\right){}_2F_2\left(\frac{4}{3}+\frac{\nu}{2},\frac{5}{6}+
\frac{\nu}{2};\frac{1}{3},\frac{2}{3};-\frac{4D\lambda_k^3}{27 \zeta^2}\right)
\right.\nonumber\\
&\ &\qquad\qquad\qquad {} - \frac{D^{1/3}\lambda_k}{|\zeta|^{2/3}} \Gamma\left(
\frac{7}{3}+\nu\right)\sin\left(\pi\frac{4+3\nu}{3}\right){}_2F_2\left(\frac{7}{6}+\frac{\nu}{2},\frac{5}{3}+\frac{\nu}{2};\frac{2}{3},\frac{4}{3};-\frac{4D\lambda
_k^3}{27 \zeta^2}\right) \nonumber\\
&\ &\qquad\qquad\qquad \left. {} + \frac{1}{2}\left(\frac{D^{1/3}\lambda_k }{
|\zeta|^{2/3}}\right)^2 \Gamma\left(3+\nu\right)\sin\left(\pi\nu\right){}_2F_2\left(2
+\frac{\nu}{2},\frac{3}{2}+\frac{\nu}{2};\frac{4}{3},\frac{5}{3};-\frac{4D\lambda_k^3}{27 \zeta^2}\right)\right].
\label{eqBE}
\end{eqnarray}
Here \eq{{}_2F_2(\cdot)} is a  hypergeometric function [SM2]. 
Eq. \eqref{eqBE} is called the Bessel distribution, and it is plotted in Fig.  \ref{fig0\iffalse BesselMeanderKesBar\fi}. The method used for finding these areal distributions in \cite{barkai2014area,kessler2014distribution}, employed an eigenfunction expansion of the solution to the Feynman-Kac formula  \cite{barkai2014area}.
This formula is used for finding the distributions of functionals of a stochastic path, in our case this is \eq{\chi_i=\int_{t_{i-1}}^{t_{i-1}+\tau_i}{v(t')\Intd t'}} and  \eq{v(t')} is the velocity trajectory corresponding to the Langevin Eq. \eq{\dot{v}=-1/v+\sqrt{2D}\Gamma(t)}. Note that we use the large \eq{v} behavior
of \eq{F(v)= - v/(1+v^2)} (the  Sisyphus cooling force in dimensionless units, as explained),  physically this works well since we are interested in the far tails of the spatial density of the particle packet, where small scale 
velocities are unimportant. 
The summation in Eqs. (\ref{eqBE},\ref{eqBM}) is performed over \eq{k} modes, where \eq{\lambda_k} are the eigenvalues of the time independent Schr\"{o}edinger-like 
equation \cite{barkai2014area}, \eq{d_k} is the normalization of the \eq{k}th eigenfunction  \cite{barkai2014area}. 
These parameters are found numerically; the method is explained in detail in \cite{barkai2014area}. Asymptotic analytic approximations and further discussion about the meaning of these parameters appear in \cite{kessler2014distribution}. 
Similarly, the distribution of the area under a Bessel meander is 
\begin{eqnarray}
B_{M}(\zeta) & =-\frac{\Gamma\left(1+\alpha\right)}{2\pi |\zeta|}\left(\frac{4^{3/2} D^{1/2}}{|\zeta|}\right)^{\nu}\sum_{k}{d_{k}}a_{k}\left[\Gamma\left(1+\nu\right)\sin\left(\pi\nu\right){}_{2}F_{2}\left(\frac{\nu}{2}+1,\frac{\nu}{2}+\frac{1}{2};\frac{1}{3},\frac{2}{3};\frac{-4D\lambda_{k}^{3}}{27\zeta^{2}}\right)\right.\nonumber \\
 & -\left(\frac{D^{1/3}\lambda_{k}}{|\zeta|^{2/3}}\right)\Gamma\left(\frac{5}{3}+\nu\right)\sin\left(\pi\frac{2+3\nu}{3}\right){}_{2}F_{2}\left(\frac{\nu}{2}+\frac{4}{3},\frac{\nu}{2}+\frac{5}{6};\frac{2}{3},\frac{4}{3};\frac{-4D\lambda_{k}^{3}}{27\zeta^{2}}\right)\nonumber \\
 & \left.+\frac{1}{2}\left(\frac{D{1/3}\lambda_{k}}{|\zeta|^{2/3}}\right)^{2}\Gamma\left(\frac{7}{3}+\nu\right)\sin\left(\pi\frac{4+3\nu}{3}\right){}_{2}F_{2}\left(\frac{\nu}{2}+\frac{7}{6},\frac{\nu}{2}+\frac{5}{3};\frac{4}{3},\frac{5}{3};\frac{-4D\lambda_{k}^{3}}{27\zeta^{2}}\right)\right].
\label{eqBM}
\end{eqnarray} 
This distribution is also plotted in Fig. \ref{fig0}. The numerical parameter, \eq{a_k} is evaluated by integration of the eigenfunction over \eq{v^{1/(2D)}} \cite{barkai2014area}. 
 In Table \ref{tablel:eigenvalues}, we provide \eq{10} values, for example, for  \eq{\lambda_k,d_k} and \eq{a_k}, with \eq{D=0.4} and \eq{D=0.5}, which are sufficient for a good approximation of the distributions.
\\ 
\\
\begin{table}[h]
\centering 
\begin{tabular}{c c c c c c c c c c c c } 
 \hline\hline 
\eq{D}&\eq{k\rightarrow} &$1$&$2$&$3$&$4$&$5$&$6$&$7$&$8$&$9$&$10$\\
[0.5ex] %
\hline 
$0.4$&$\lambda_k$&$3.593$&$5.075$&$6.3855
$&$7.5581$&$8.657$&$9.691$&$10.673$&$11.613$&$12.515$&$13.387$ \\

$\qquad$ &$d_k$&$0.4159$&$0.57$&$0.682 $&$0.773$&$0.851$&$0.920$&$0.983$&$1.042$&$1.095$&$1.145$ \\

$\qquad$&$a_k$&$0.849$&$0.314 $&$0.535$&$0.307$&$0.441$&$0.295$&$0.3915$&$0.284$&$0.359$&$0.274$  \\
\hline 
$0.5$&$\lambda_k$&$3.37$&$4.89 $&$6.21$&$7.41$&$8.52$&$9.56$&$10.55$&$11.5$&$12.4$&$13.28$ \\

$\qquad$ &$d_k$&$0.521$&$0.675$&$0.78 $&$0.879$&$0.94$&$0.997$&$1.076$&$1.102$&$1.151$&$1.221$ \\

$\qquad$&$a_k$&$1.040$&$0.239 $&$0.604$&$0.244$&$0.475$&$0.225$&$0.414$&$0.215$&$0.365$&$0.217$  \\
  
\hline 
\end{tabular} 
\caption{The first $10$ numeric coefficients, $\lambda_k$, $d_k$ and $a_k$, required for plotting the theoretical PDFs, Eqs. (\ref{eqBE},\ref{eqBM}), with \eq{D=0.4} and \eq{0.5}. These coefficients were calculated using the method explained in \cite{barkai2014area}. As explained, these values are also  required for plotting the ICD, \eq{\mathcal{I}(z)}, Eq. (\eq{12}) in the main text. For this purpose as well, we found that \eq{10} \eq{k}-modes are sufficient.}
\label{tablel:eigenvalues} 
\end{table}  

\subsection{D. Derivation of the moments}
We derive the moments \eq{\langle x^{2m}\rangle}, for \eq{m=1,2,...}, presented in Eq. (\eq{7}) in the main text. Our starting point is Eq. \eqref{eqMW14} and the areal distribution of the excursions. Applying Fourier \eq{x\rightarrow k} and Laplace \eq{t\rightarrow u} transforms to Eq. \eqref{AppModMW:DeltaM}, using Eq. \eqref{APPMeander}, we write 
\EQ{\hat\Phi_E \left(k,u\right) = \int_{-\infty}^\infty{d\chi\int_0^\infty{d\tau e^{-u\tau+ik\chi} g \left(\tau\right)	\frac{1}{\tau^{3/2}}B_E\left(\frac{\chi}{\tau^{3/2}}\right)}}.}{LFTPhiDist} Note that the distribution \eq{g(\tau)} is given by the solution of a standard first passage time problem in \cite{marksteiner1996anomalous}. Asymptotically,       \begin{equation} 
g(\tau) \approx g_*\tau ^{-1-3\nu/2}, \quad\mbox{for}\quad \tau\gg1
\label{eq:g_t_w_t_cold}
\end{equation} 	
where \eq{\nu=1/(3D)+1/3} and (see \cite{barkai2014area})
\begin{equation} 
  \frac{g_*}{\langle\tau\rangle}=\frac{2D}{\sqrt{\pi}}\frac{(1+D)\Gamma(\frac{1}{2D})}{(4D)^{\frac{1+D}{2D}}\Gamma(\frac{1-D}{2D})\Gamma(\frac{1+D}{2D})}. 
\label{eq:g_over_tau} 
\end{equation} 
Expanding the exponent \eq{e^{ik\chi}} in Eq. \eqref{LFTPhiDist} as a Taylor series for small \eq{k}, using Eq. \eqref{eq:g_over_tau}, while separating out the \eq{n=0}'th term, we obtain  
\begin{equation} 
\hat{\Phi}_E(k,u) = 
\hat g(u)+\sum_{m=1}^\infty{\frac{(ik)^m}{m!}\langle\zeta^m\rangle_E {g^*}\Gamma(\frac{3m}{2}-\frac{3\nu}{2})u^{-\frac{3m}{2}+\frac{3\nu}{2}}}, 
\label{eq5}
\end{equation} 
where \eq{\hat g(u)=\int_0^\infty g(\tau)\exp(-u\tau)\Intd \tau} and \EQ{\langle\zeta^m\rangle_E=\int_{-\infty}^\infty{\zeta^m B_E(\zeta)\Intd{\zeta}}.}{xinE} Here, we changed variables to \eq{\zeta=\chi/\tau^{3/2}}. Notice that \eq{B(\zeta)} is symmetric, hence its odd moments are zero. Using the equivalent procedure for $\hat\psi_M\left(k,u\right)$ with $B_M(\zeta)$, and using \EQ{\hat w(u)=\int_0^\infty{w(\tau)\exp(-u\tau)\Intd \tau}=[1-\hat g(u)]/u,}{Wu} for the survival probability \eq{w(\tau)} (defined in Sec. B.), we rewrite Eq. (\ref{eqMW14}) as 
\begin{equation} 
\hat {P}_u\left(k\right) = \frac {1}{u}\frac {1+\sum_{m=1}^\infty{\frac{(-1)^{m}}{\hat w(u)}\frac{2{g^*}}{3\nu}\Gamma({3m}-3\nu/2+1)\frac{1}{(2m)!}\langle\zeta^{2m}\rangle_Mk^{2m} u^{-{3m}+3\nu/2-1}}}{1-\sum_{m=1}^\infty{\frac{(-1)^{m}}{1-\hat g(u)}{g^*}\Gamma({3m}-3\nu/2)\frac{1}{(2m)!}\langle\zeta^{2m}\rangle_Ek^{2m} u^{-{3m}+3\nu/2}}}. 
\label{eq:Dist} 
\end{equation} 

For \eq{2/3<\nu<2} (recall, we derive our main results in the range \eq{0<D<1}), the average $\langle\tau\rangle$ is finite, therefore from Eqs. (\ref{eqMW14},\ref{eq:g_t_w_t_cold}), the Laplace-transforms of \eq{g(\tau)} and \eq{w(\tau)}, in the \eq{u\rightarrow0} limit are 
\begin{equation} 
\hat g(u)\approx 1-u\langle\tau\rangle, \qquad\qquad 
\hat w(u)\approx \langle\tau\rangle. 
\label{eq8} 
\end{equation}
By using the relation \eq{\langle x^{2m}\rangle=(-1)^m\left[\Intd^{2m}/\Intd k^{2m} \hat{P}_u(k)\right] \iffalse\bigg\rvert\fi|_{k=0}}, we derive the moments in Eq. (\eq{6}), in the main text, in the long time limit.  
\\ 
\\
\footnotesize{(SM1) R. Metzler and J. Klafter, Physics reports \textbf{339}, 1 (2000)}\\ 
\footnotesize{(SM2) M. Abramowitz and I. A Stegun, \em{Handbook of Mathematical functions: with formulas, graphs, and mathematical tables} (Courier Dover Publications, 1972).}

\end{document}